\documentclass[screen,acmsmall]{acmart}

\usepackage{array}
\usepackage{fontawesome5}
\usepackage{xcolor}

\newcommand{\system}{SpeechCap}

\newcommand{\altcolor}[1]{{#1}}

\AtBeginDocument{%
  }

\setcopyright{cc}

\copyrightyear{2025}
\acmYear{2025}

\acmISBN{978-1-4503-XXXX-X/2018/06}

\begin{document}

\title{SpeechCap: Leveraging Playful Impact Captions to Facilitate Interpersonal Communication in Social Virtual Reality}
\renewcommand{\shorttitle}{SpeechCap}

\author{Yu Zhang}
\email{yui.zhang@my.cityu.edu.hk}
\orcid{0000-0002-8574-111X}
\affiliation{%
  \department{Department of Computer Science}
  \institution{City University of Hong Kong}
  \city{Hong Kong SAR}
  \country{China}
}

\author{Yi Wen}
\email{cyberwenyi2357@tamu.edu}
\affiliation{%
  \institution{Texas A\&M University}
  \city{College Station}
  \state{Texas}
  \country{USA}
}
\affiliation{%
  \department{School of Creative Media}
  \institution{City University of Hong Kong}
  \city{Hong Kong SAR}
  \country{China}
}

\author{Siying HU}
\email{sying.ch1026@gmail.com}
\affiliation{%
  \department{Department of Computer Science}
  \institution{City University of Hong Kong}
  \city{Hong Kong SAR}
  \country{China}
}

\author{Zhicong Lu}
\email{zlu6@gmu.edu}
\affiliation{%
  \institution{George Mason University}
  \city{Fairfax}
  \state{Virginia}
  \country{USA}
}

\renewcommand{\shortauthors}{Zhang et al.}

\begin{abstract}
Social Virtual Reality (VR) emerges as a promising platform which affords immersive, interactive, and engaging mechanisms for collaborative activities in virtual spaces. 
However, interpersonal communication in social VR is still limited with existing mediums and channels.
To bridge the gap, we propose a novel method for mediating real-time conversations in social VR, which leverages \textit{impact captions}, a type of typographic visual effect widely used in videos, to encode both verbal and non-verbal information.
We first investigated the design space of impact captions by content analysis and a co-design session with four experts.
We then implemented \system{}, a proof-of-concept system with which users can communicate with each other using speech-driven impact captions in VR.
Through a user study (N=14), we evaluated the effectiveness of the visual and interaction design of speech-driven impact captions, highlighting their strengths in the interactivity and integrating verbal and non-verbal information in communication mediums.
Finally, we discussed our main findings regarding visual rhetoric, interactivity, and ambiguity, and further provided design implications for facilitating interpersonal communication in social VR.
\end{abstract}



\begin{teaserfigure}
  \includegraphics[width=\textwidth]{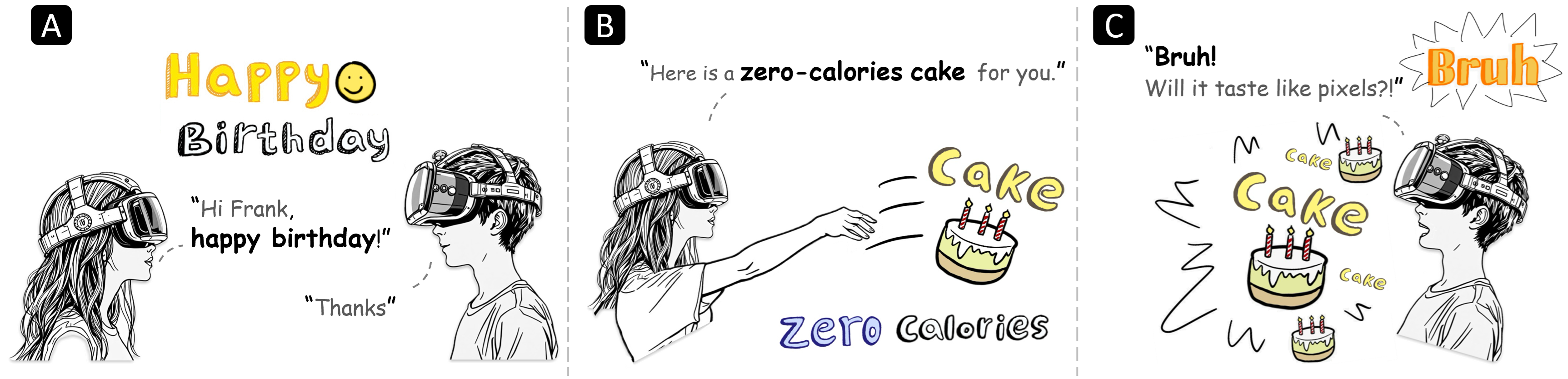}
    \caption{
    SpeechCap showcases a novel approach to mediate voice conversations in social VR by converting speech into playful impact captions.
      (A) From real-time speech, SpeechCap generates impact captions with customized visual design that conveys both verbal and non-verbal information. The caption ``happy'' is in a warm and bright color for the positive emotions, and a ``smiling'' emoji further highlight the pleasant and relaxing atmosphere.
      (B) Users can play with impact captions like real-world objects, e.g., throwing a ``cake''.
      (C) Once the flying ``cake'' crashes on another user, it will explode like a firework.
    }
  \Description{.}
  \label{fig:teaser}
\end{teaserfigure}


\maketitle

\section{Introduction}
With the increasing availability of consumer-level virtual reality head-mounted displays (HMDs) ~\cite{anthes2016state, li2021social}, Social Virtual Reality (VR) emerges as a promising platform that brings immersive, interactive, and engaging experiences to users for a wide range of collaborative activities. 
Social VR now supports remote conferencing ~\cite{abdullah2021videoconference}, immersive learning ~\cite{thanyadit2022xr, peng2021exploring}, healthcare ~\cite{li2020designing, udapola2022social}, collaborative design ~\cite{mei2021cakevr}, music composing ~\cite{men2022supporting} and even daily activities such as sleeping ~\cite{maloney2020falling} and drinking ~\cite{chen2024drink}.
Among these diverse application areas, social VR helps users overcome physical dispersion, enabling them to communicate and collaborate in the same virtual environment ~\cite{li2021social, palmer1995interpersonal} by offering channels for both verbal and non-verbal communication ~\cite{dzardanova2022virtual, mcveigh2021case, wei2022communication}.
Specifically, text messaging and voice chat are two channels commonly used for verbal communication. 
And pictures ~\cite{li2019measuring}, avatars ~\cite{fu2023mirror, baker2021avatar, kolesnichenko2019understanding, freeman2021body}, and embodied interactions ~\cite{maloney2020talking, wieland2022non} are channels that can convey non-verbal information in social VR.

Yet, these existing communication channels in social VR are preliminary, as they are either borrowed from traditional media (e.g., texts, voices, and pictures) lacking adaptations to the nature of social VR ~\cite{montoya2023wordsphere, rzayev2021reading, dzardanova2022virtual, hsieh2020bridging}, or difficult to be reproduced very well by current VR technologies (e.g., gestures, facial expressions, embodied actions) ~\cite{sykownik2023vr, tanenbaum2020make, wei2022communication}.
Specifically, texts as a communication medium in VR still suffer from difficulties in user inputting ~\cite{montoya2023wordsphere} and readability issues in presentations ~\cite{rzayev2021reading, hsieh2020bridging}. 
Voice-based conversations in VR can easily be disturbed by sound intrusions from the environment ~\cite{akselrad2023body}. 
Furthermore, non-verbal communication mediums, such as gestures and facial expressions, cannot be well-simulated with the avatars in current VR applications ~\cite{baker2021avatar, aseeri2021influence, freeman2021body}, and users may find it hard to achieve specific actions or movements using controllers with joysticks and buttons ~\cite{sykownik2023vr, tanenbaum2020make, li2019measuring}.

Except for social VR, previous research in HCI and CSCW has explored ways to enhance interpersonal communication in different scenarios. Those scenarios include video conferencing ~\cite{liu2023visual, xia2023crosstalk}, augmented reality (AR) ~\cite{lee2023exploring, liao2022realitytalk, leong2022wemoji}, and text messaging interfaces ~\cite{aoki2022emoballoon}. 
These different scenarios demonstrate the power of the integration of both verbal and non-verbal cues in the mediums to support interpersonal communication. 
On the other hand, researchers have also envisioned the potential of social VR in providing situational awareness and enabling new social interactions with novel communication mediums that are unattainable in traditional computer-mediated communication (CMC) platforms ~\cite{mcveigh2021case, mcveigh2022beyond}. 

On top of the prior work, we proposed a new approach to support interpersonal communication in social VR. Our approach is inspired by \textit{Impact Captions}, a type of expressive typographic 2D visual effect with engaging visual elements and animated motions. Impact captions are traditionally used in entertaining television shows and online videos ~\cite{o2010japanese, sasamoto2014impact} to attract viewers' attention and create engaging watching experiences ~\cite{sasamoto2021hookability}. 
Particularly, we introduced impact captions as a communication medium to facilitate interpersonal communication and interactions in social VR. In our design, impact captions encode both verbal and non-verbal information extracted from real-time speech; they also offer interactivity that allows users to intentionally play with the captions for communicative purposes.

To implement and evaluate our impact-caption-inspired approach, we first explored the visual and interaction design space of impact captions through a content analysis on TV show videos.
Based on the design space, we held a co-design session with 4 experts (i.e., HCI researchers) to further derive specific design goals for building a system to demonstrate our idea.
Following the findings so far, we developed \system{}, a proof-of-concept system that takes real-time speech as input to generate interactive impact captions with customized visual appearances in a multi-user VR environment.
Using \system{}, we conducted an in-lab study with 14 participants to assess impact captions as a medium for supporting interpersonal communication in social VR. 

The user study results highlighted the integration of verbal and non-verbal information offered by impact captions, indicating the design can make social VR communication experiences not only clear but also engaging, while the risks of miscommunication caused by ambiguity inherent from the complex visual design were noticed.
Moreover, we found that impact captions can also facilitate interactions between users beyond enhancing speech conversations in social VR.
Furthermore, we illustrated three application scenarios to demonstrate the generalizability of our impact-caption-inspired approach. 
Finally, this research ends up with discussions on the findings from the user study, in which we provided insights and implications for future work and explained the limitations and possible improvements on the scalability of the current \system{} system.

In summary, this research contributes:
\altcolor{
(i) A design space of impact captions for supporting interpersonal communication in social VR;
}
(ii) A proof-of-concept system, \system{}, that enables users to have real-time voice-based conversations with the support of interactive impact captions in social VR;
(iii) An evaluation study of \system{} that demonstrates the effectiveness of our impact-caption-inspired approach for communication, highlighting the value of employing creative mediums for supporting interpersonal communication in social VR.

\section{Related Work}
Starting from the background of interpersonal communication in social virtual reality (VR), we highlight the needs for better ways to support communication with both verbal and non-verbal information, and then introduce previous research systems to augment computer-mediated communication (CMC), which inspired our work. Finally, we introduce impact captions and captioning systems for communication purposes in HCI, showing our motivation for this research.

\subsection{Interpersonal Communication in Social Virtual Reality}
Social Virtual Reality (VR) refers to immersive virtual spaces where multiple users can interact synchronously through VR head-mounted displays (HMDs) ~\cite{mcveigh2019shaping, freeman2021body}. It provides high-fidelity virtual presence to facilitate various forms of interpersonal communication with both verbal and non-verbal cues over other computer-mediated communication channels ~\cite{yassien2020design, mcveigh2021case}. 

In an immersive and synchronous environment, social VR can effectively support collaborative tasks and promote social interactions in a wide range of application scenarios, including collaborative prototyping ~\cite{mei2021cakevr}, healthcare and treatment ~\cite{li2020designing, udapola2022social}, immersive learning ~\cite{thanyadit2022xr, peng2021exploring, jensen2018review}, intimate relationship building ~\cite{wang2023designing, freeman2021hugging}, and inter-generational communication ~\cite{shen2024legacysphere, du2024ai, wei2023bridging, baker2019exploring}.

Although social VR has been increasingly applied in various domains, it is still far from fulfilling users’ social and emotional needs as a digital space ~\cite{wei2022communication, tanenbaum2020make, maloney2020talking}. Previous research identified several limitations regarding interpersonal communication in social VR.
A known issue is that current social VR applications do not allow users to easily express emotions (e.g., mood and excitement) and intentions through non-verbal cues, which are common in real-world conversations ~\cite{maloney2020talking, sykownik2023vr, wu2023interactions}, because virtual avatars reduce or even filter out several important non-verbal signals such as facial expressions and body language ~\cite{baker2021avatar, freeman2021body, zhang2022s, lee2022understanding, fu2023mirror}. As an extreme case, imperfect avatars with realism will lead to the uncanny valley effect, bringing discomfortable social experiences to users ~\cite{latoschik2017effect, kyrlitsias2022social}. 
Besides the lack of expressiveness, voice-based communication, a common method in social VR, faces challenges in coordinating speakers in multi-user virtual spaces ~\cite{yan2023conespeech}.

Recent research proposed several computational methods to improve interpersonal communication experiences in social VR, such as generating realistic avatars that could capture users' facial expressions and body movements using deep learning models ~\cite{van2022deep}, adding haptic technologies to facilitate co-presence to allow affective communication ~\cite{fermoselle2020let, ahmed2016reach}, and simulating spatial sound effects to enhance the convenience and flexibility of voice-based communication ~\cite{yan2023conespeech}. 
In summary, all of these efforts aim to make the interpersonal communication in social VR close to the situations of the realistic world. 

However, people appreciate social VR not because it can simulate the real world. Instead, a significant reason is that 
the unique affordances of social VR can offer users immersive and unrealistic social experiences ~\cite{freeman2021body, maloney2020talking}. In other words, social VR can augment social signaling and unlock new social interactions that are unattainable in the realistic world ~\cite{mcveigh2022beyond, mcveigh2021case}.
At present, the potential of virtual reality to create unrealistic experiences for enhancing interpersonal communication for socialization needs is still under-explored. This leaves us opportunities to dive deep into the potentials of virtual reality and build novel tools with the idea of ``superpowers'' ~\cite{mcveigh2022beyond, mcveigh2021case} to support interpersonal communication in social VR.

\subsection{Combining Verbal and Non-verbal Information to Augment Communication}
Both verbal and non-verbal information is crucial in interpersonal communication in social VR ~\cite{palmer1995interpersonal}.
Besides verbal cues that mainly convey semantic meanings, non-verbal cues convey emotions ~\cite{liebman2016s, luo2024emotion}, enhance understanding ~\cite{aburumman2022nonverbal}, regulate interactions ~\cite{maloney2020talking}, reflect cultural norms ~\cite{freeman2021hugging}, and overall enrich communication experiences, whether in offline face-to-face scenarios or online digital spaces.

Recent research in the HCI and CSCW fields has explored ways to involve non-verbal signals in verbal-centric communications (e.g., live presentations) to facilitate interpersonal communication and build social connections through computational approaches ~\cite{liao2022realitytalk, liu2023visual, cao2024elastica, an2024emowear, chen2021bubble, aoki2022emoballoon, choi2019emotype}.
A thread of research aims to provide visual aids for computer-mediated spoken presentations. ``RealityTalk'' ~\cite{liao2022realitytalk} explores a speech-driven approach that allows users to give a speech with predefined visual elements while mentioning particular key words. ``Visual Captions'' ~\cite{liu2023visual} and ``CrossTalk'' ~\cite{xia2023crosstalk} further automate the process by introducing machine-learning-based models to predict users' intentions and automatically provide potential visual elements while processing speech input. ``Elastica'' ~\cite{cao2024elastica} further enables adaptive animations on the associated visuals to effectively enhance the expressiveness of presentations.

Beyond visual aids, enhancing emotional communication is another important dimension explored by previous research. For voice messaging interfaces, ``Emowear'' ~\cite{an2024emowear} introduced the concept of ``Emotional Teasers,'' a collection of animated emoticons designed to show the emotional tone of upcoming voice messages for smartwatch users.
For text-centered communication mediums such as text messaging interfaces, previous research has explored how the color and shape of speech bubbles ~\cite{chen2021bubble, aoki2022emoballoon}, and the typeface of textual content ~\cite{choi2019emotype, de2023visualization} can be purposefully designed and computationally generated to convey the speaker's moods in conversations. 

However, still less research explores how communication mediums in social VR can borrow the ideas from other CMC forms to enhance the interpersonal communication experiences, especially with the considerations of integrating both verbal and non-verbal information.
So far, previous research revealed that non-verbal cues play an important and unique role in interpersonal communication in social VR.
In unrealistic virtual spaces, users tend to engage in social interactions with more active and bold non-verbal behaviors ~\cite{maloney2020falling, chen2024drink}. 
This phenomenon reflects a ``proteus effect'' in which users intentionally manipulate virtual avatars to express themselves beyond pure verbal-language-based communication ~\cite{maloney2020talking}. 
With more application scenarios and novel user activities emerging in social VR, strong demands for tools for facilitating interpersonal communication and interactions arise ~\cite{tanenbaum2020make}.

\subsection{Impact Captions: A Typographic-Centered Visual Design Beyond Pure Verbal Communication}
Impact captions refer to a type of typographic-centered visual effect that is prevalent in TV shows and online videos to engage viewers ~\cite{sasamoto2014impact}. 
Traditionally, impact captions are used as a supplementary channel to provide information outside the screen (e.g., commentary messages from the TV show's editor ~\cite{o2010japanese}) or to visualize implicit non-verbal information within frames (e.g., characters’ moods or environmental sound effects ~\cite{wang2016visualizing}).
Unlike visual cues in previous work for improving interpersonal communication ~\cite{liao2022realitytalk, liu2023visual, cao2024elastica, an2024emowear, chen2021bubble, aoki2022emoballoon, choi2019emotype}, impact captions always use text as the main visual component that conveys verbal information and apply artistic modifications with additional visual decorations to textual elements to convey non-verbal cues simultaneously. The integration of both verbal and non-verbal information in the visual design of impact captions makes the captions a powerful medium that can provide rich semantics, evoke emotional reactions in viewers, and provoke thoughts ~\cite{o2010japanese, sasamoto2021hookability, chow2023impact}.

HCI researchers have noticed the potential of captioning mechanisms in supporting communication and have built captioning systems to help people with deaf and hard-of-hearing (DHH) situations perceive in-depth information in videos ~\cite{kim2023visible, de2023visualization, bragg2017designing, wang2016visualizing, seto2010subtitle}. In these cases, modifications on textual elements (e.g., irregular typefaces, coloring, and introducing dynamic visual effects) of subtitles have been proved to be effective in conveying semantic information of both verbal words and implicit non-verbal content of videos ~\cite{kim2023visible, de2023visualization, seto2010subtitle}. Yet, their explorations have not reached the scope of immersive media, such as VR.
When considering textual captions in VR, another line of research indicates that the reading experience of texts in VR varies depending on the displaying conditions ~\cite{rzayev2021reading}, and prolonged reading would reduce the comfort of using VR ~\cite{ubur2024easycaption}. 
Yet, it is also unknown whether typographic-centered impact captions can effectively mediate communication in social VR.

To fill the research gap on tools for interpersonal communication in social VR and to respond to the needs of social VR users, this work explores how impact captions, as a type of elegant typographic design, can be applied to mediate and facilitate interpersonal communication in social VR with the inspirations from previous research on augmenting computer-mediated communication using visual cues and captioning mechanisms.
By investigating a design space for using impact captions as a communication tool and conducting a study with the proof-of-concept system, we demonstrate a concrete example envisioning the potentials of social VR in interpersonal communication and socialization.

\section{Design Space of Impact Captions}
\label{section_design_space}
In order to introduce impact captions into social VR as a medium for interpersonal communication, we explored the design space of impact captions through a content analysis on the TV variety show videos.
Our design space consists of the visual design (\autoref{fig:ds_visual_elements}) and the interaction design (\autoref{fig:ds_interactions}) of impact captions. 
Furthermore, the visual design space separately considered textual elements and non-textual elements.

\subsection{Methods: Analyzing Videos with Impact Captions}
Three researchers conducted the content analysis collaboratively on a collection of videos of TV variety show, in which impact captions are commonly used, to explore the design of impact captions.



\subsubsection{Data Collection}
We collected relevant videos using a two-stage approach.
Firstly, we followed the top TV series rankings (i.e., IMDb\footnote{https://www.imdb.com/?ref\_=nv\_home}, TV Time\footnote{https://www.tvtime.com/}, and Douban\footnote{https://m.douban.com/tv/tvshow}) to select TV shows where impact captions are widely used. 
Specifically, we concentrated on three representative variety show series: ``Arashi'' (from Japan), ``Running Man'' (from South Korea), and ``Who is the Murderer'' (from China). For each show, we reviewed one randomly-selected episode.
Secondly, we enlarged the collection by including videos found from YouTube and Bilibili, which are two popular online video-sharing platforms. 
On both platforms, we used the keywords ``variety show with captions'' and ``popular variety shows'' (we translated the keywords into Chinese version on Bilibili) to search for videos using a newly registered account. From the raw results of each keyword, we took the top-30 videos and filtered out irrelevant or redundant videos. Finally, 46 sample videos were included.


\subsubsection{Content Analysis}
\altcolor{
Three authors collaboratively conducted content analysis on the video collection to build the design space. 
At the beginning, three authors started from a subset (N=10) of videos, 3 from the three popular TV shows and 7 randomly-selected online videos, to individually code the visual elements and behaviors of the impact captions in the videos, with the aim of determining initial dimensions. 
Through a round of discussion and merging on their initial insights, three authors ultimately determined the dimensions in the design space, including textual visual dimensions (i.e., typeface, size and color), non-textual visual dimensions (i.e., ornament, speech bubble and emoji), and interactions (i.e., physicalization, motion, embodied interaction).
}

\altcolor{
Using the determined dimensions as a codebook, each of the three authors took 13 videos (i.e., one third of the rest videos) to review. During the process, the reviewers also had multiple rounds of discussions on the possible values of some dimensions (e.g., spiky bubbles in speech bubble).
As a result, we found that all of the 49 (i.e., 3 from the TV shows plus 46 online) videos manipulated both textual and non-textual visual dimensions to design impact captions. For the dimensions under the interaction category, 29 videos significant used motions, 13 videos had physicalization, and 15 made impact captions to interplay with human characters.}


\subsection{The Visual Design of Textual Elements}
As the primary visual component, the textual elements of impact captions follow the principles of general typographic design ~\cite{carter2011typographic}, in which \emph{typeface}, \emph{size}, and \emph{color} are three main visual dimensions.


\begin{figure}[htb]
    \includegraphics[width=12cm]{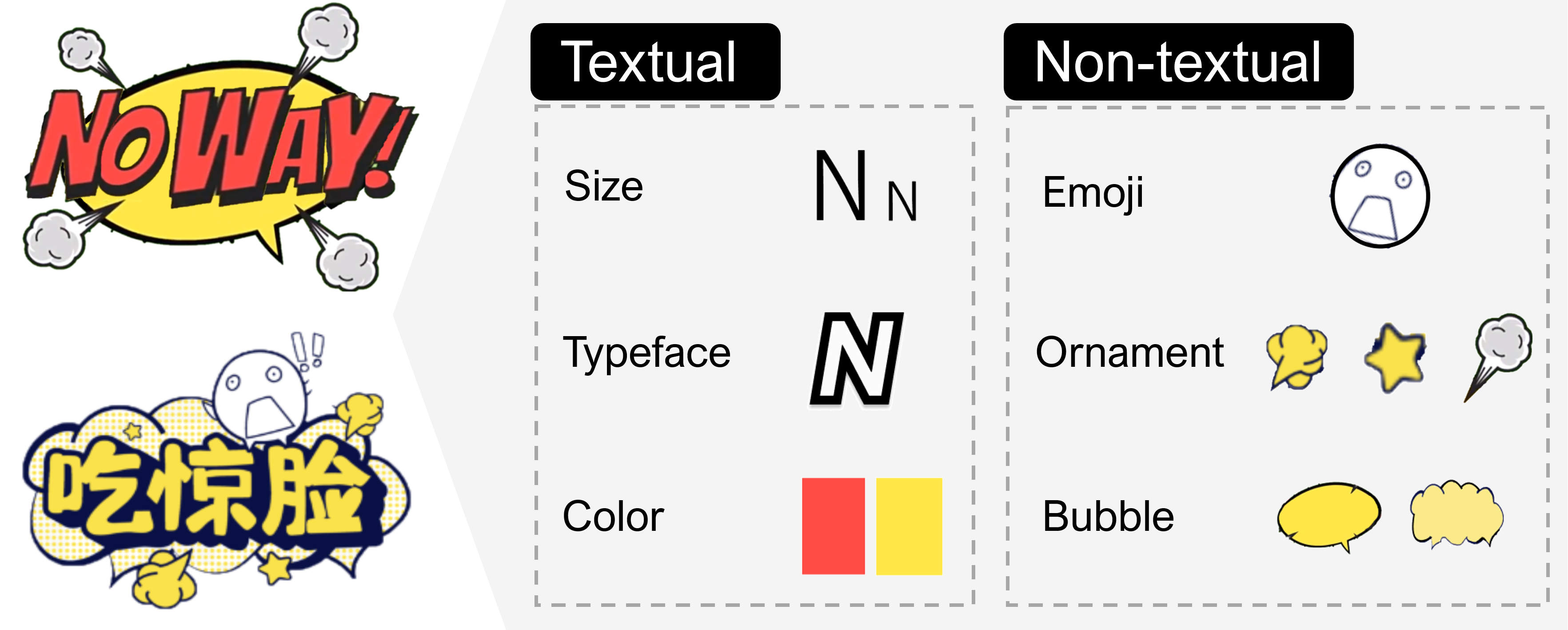}
    \caption{The Visual Design Space of Impact Captions.
    Two impact captions, a ``No Way'' and a ``Shocked Face'' (translated from Chinese), demonstrate the \textbf{textual} and \textbf{non-textual} elements with relevant dimensions that form the visual design space. 
    For textual elements, the typeface, the color, and the size of texts are three major dimensions. 
    Non-textual elements include emoji, ornament, and speech bubble.}
    \Description{.}
    \label{fig:ds_visual_elements}
\end{figure}

\subsubsection{Typeface}
Typeface denotes the fundamental visual style of each letters in a piece of text ~\cite{carter2011typographic}. It can affect the perceptions and emotional responses of the readers ~\cite{bianchi2021emotional, amare2012seeing}. 
Impact captions widely utilize typeface to make impressions for the audiences ~\cite{sasamoto2021hookability}. 
Although there is no strict standard that defines which typeface should be used under which circumstance, we found that the regularity of strokes can relate to the content of impact captions. Specifically, artistic typefaces with irregular strokes are usually used to highlight characters' moods or to emphasize surprising situations, while formal and regular typefaces (e.g., serif fonts) are used for presenting neutral and factual information.



\subsubsection{Size}
The size of the textual elements is a powerful dimension used for several purposes.
First, impact captions with changeable sizes attract attention, making the particular text content stand out on the screen to engage viewers.
Second, larger texts helps enhance the readability and accessibility of video content.
Third, size can convey liveliness atmosphere and the excitement of characters.
Additionally, size can also encode some highly abstract semantic meanings, such as strength, effectiveness, and severity.

\subsubsection{Color}
\label{sec_space_color}
Similar to size, color can be vibrant and eye-catching to attract the audience's attention and highlight important information. 
In addition, color of the textual elements of impact captions is commonly used to imply the moods of characters in TV shows, which reflects the color-emotion associations in human perceptions ~\cite{plutchik2013theories, hanada2018correspondence}.
Although the perceptions of the emotion conveyed by the same color can vary among different people ~\cite{wilms2018color, hanada2018correspondence, chen2021bubble}, the color-emotion associations still reveal the potential of color as a dimension to convey emotional information for bi-directional interpersonal communication.



\subsection{The Visual Design of Non-textual Elements}
We identified \emph{typographic ornaments}, \emph{speech bubbles}, and \emph{emojis \& emoticons} as three types of commonly used non-textual elements of impact captions through the content analysis.

\subsubsection{Typographic ornaments}
Typographic ornaments refer to the decorative elements or symbols used to embellish and enhance the appearance of impact captions. They are typically small, non-alphabetic characters that add artistic flair and visual interest to the textual elements.
For example, the ``erupting gas'' ornaments strengthen the tone of saying ``No Way'' (\autoref{fig:ds_visual_elements}).

\subsubsection{Speech bubbles}
Speech bubbles (a.k.a. speech balloons, ornamental frames) refer to the background containers to display text inside.
In comics, speech bubbles is widely used for presenting speech or thoughts and indicating the source of the speech or thoughts with a tail pointing to characters ~\cite{cohn2013beyond}. 
As a non-textual visual element in impact captions, speech bubbles not only inherit the functionalities as they have in comics, but also convey non-verbal information such as the emotion of characters and the tone of speech through shape and color. This feature has been in other contexts like text messaging ~\cite{aoki2022emoballoon}.

\subsubsection{Emojis and Emoticons}
Emojis and Emoticons, as non-textual glyphs that represent facial expressions ~\cite{lo2008nonverbal}, are a special group of decorative elements used in impact captions to convey feelings or simulate human reactions that can not be easily expressed through texts alone.
For example, the ``amazed face'' emoji in the lower impact caption in \autoref{fig:ds_visual_elements} visually repeats the meaning of the text to enhance the impressiveness.



\subsection{The Interaction Design of Impact Captions}
\altcolor{
Shifting from video to VR, impact captions are no longer post-edited visual effects but become ``alive''---real-time rendering allows impact captions to lively interact with users, while the human-caption interactions in VR remain under-explored. 
Through the content analysis, we found that impact captions in TV shows are often designed to simulate physical properties and possess motion effects, or made to coordinate with the embodied actions of human characters. 
Inspired by these post-edited behaviors, we explored the interaction design space of impact captions for VR, considering \emph{Physicalization}, \emph{Motion}, and \emph{Embodied Interaction}.
}


\subsubsection{Physicalization}
\begin{figure*}[ht]
    \centering
    \includegraphics[width=\linewidth]{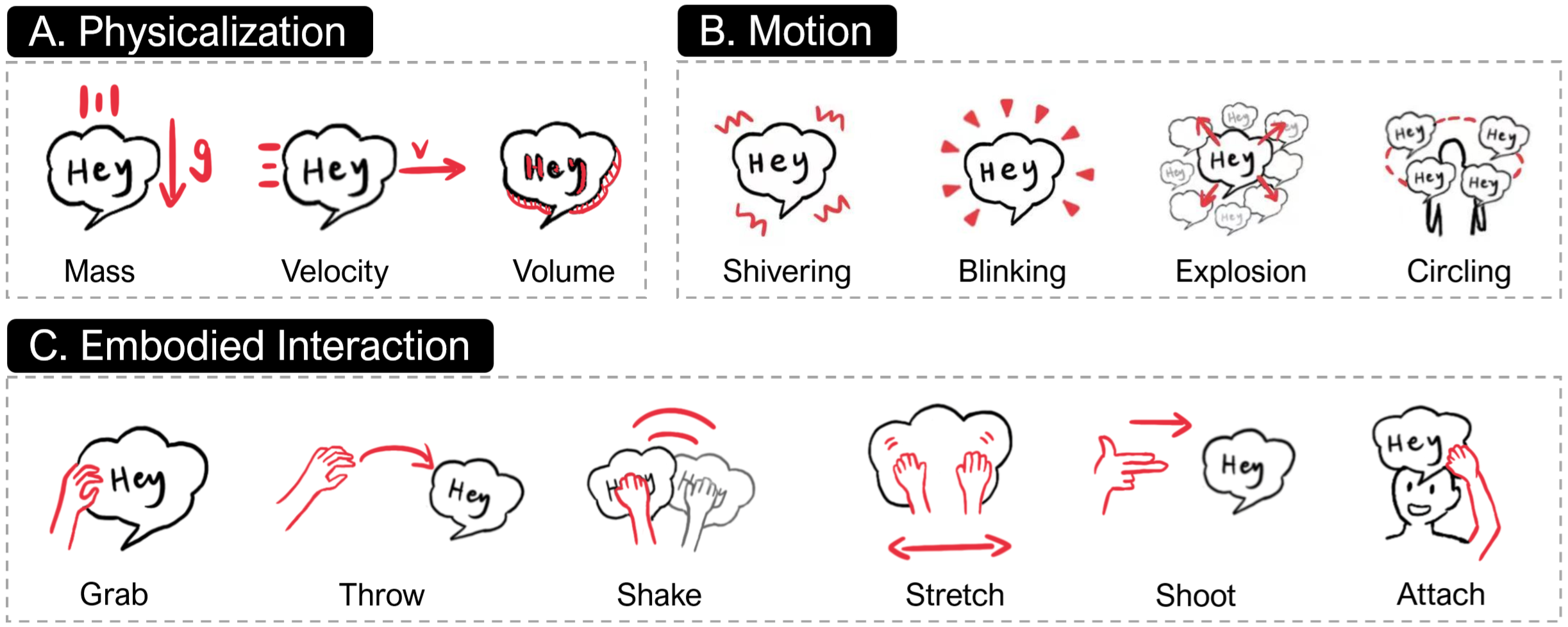}
    \caption{
        The Interaction Design of Impact Captions.
        The design space includes three dimensions: \textbf{Physicalization}, \textbf{Motion}, and \textbf{Interaction}.
        With Physicalization, impact captions can appear like physical objects to be affected by the gravity with mass, have velocity for movement, and take spaces with volume.
        With Motion, impact captions appear to be ``alive'' and be responsive to user actions.
        Interaction describes how users can play with impact captions using embodied interaction.
    }
    \Description{.}
    \label{fig:ds_interactions}
\end{figure*}

Physicalization allows virtual objects to simulate physical properties to create tangible sensations for users to enrich their interactions with digital user interfaces ~\cite{sauve2024physicalization, hornecker2006getting}.
We found that impact captions can also appear like they have physical properties in TV show videos to convey non-verbal meanings. Three commonly used physical properties are mass, velocity, and volume (\autoref{fig:ds_interactions} A). 
Mass implies the perception of weight and can be used to represent the feeling of ``light'' and ``heavy'' of abstract concepts such as the seriousness of speech. 
Velocity associates with time, indicating the sense of urgency. 
Volume relate to the size of text in impact captions but the overall volume of an impact caption can still convey the sense of ``small'' and ``big'' without words.







\subsubsection{Motion}
Motions make impact captions alive and attractive by endowing changing appearances. 
Four kinds of motions were found and included in our design space: shivering, blinking, and explosion (\autoref{fig:ds_interactions} B).
Shivering means that impact captions will swing periodically . 
Blinking makes the color of visual components (e.g., text or speech bubble) change rapidly and glowing.
Explosion is an effect where an original caption splits into several pieces of fragments or replicas spreading around.
Circling makes impact captions orbit an object or human characters.


\subsubsection{Embodied Interaction}
Embodied interaction emphasizes the integration of the human body and physical actions, such as gestures ~\cite{luo2024emotion} and bodily movements ~\cite{mueller2018experiencing}, as integral components of interaction with digital technology ~\cite{smith2018communication, kirsh2013embodied}.
Inspired by the patterns of how impact captions interplay with characters in TV shows, we include four gestures (i.e., drag, drop, stretch, and throw) and a bodily movement (i.e., attach to avatar) in our design space.
With these embodied interactions as triggers, users can intuitively manipulate and play with impact captions in VR.


\section{Co-designing an Impact-Caption-Inspired Communication Tool in Social VR}
To operationalize our impact-caption-inspired approach for interpersonal communication in social VR, we organized a co-design activity to obtain and refine the design goals for build a proof-of-concept system.

\subsection{Procedure}
The procedure consists of two sub-sessions. The first focused on brainstorming and open-ended discussion, and the second aimed at paper prototyping for the proof-of-concept system.

\subsubsection{Participants}
Besides the four authors of this paper, we recruited four extra HCI researchers from our university to this session. During the procedure, the first author played the role of instructor to host the sessions and also participated in the activities to contribute ideas. 
The eight contributors are either developers or designers of digital products. They all have rich experiences in VR gaming and social VR (e.g., VRChat\footnote{https://hello.vrchat.com/}).

\subsubsection{Sub-session 1: Brainstorming and Discussion}
In the first sub-session, the objective was to understand the scenarios, needs, and purposes of users' interpersonal communication behaviors in social VR. 
This session was followed by an initial brainstorming and discussion activity in which the participants shared their own experiences of interpersonal communication in VR to identify the potential pain points and possible improvements. 
The instructor then presented several examples of impact captions (sourced from the sample videos for investigating the design space) to illustrate the concept and potential usages. 
With the inspiration of impact captions, next the participants presented their views on how the captions can be applied to improve the communication experience in social VR.
The results of this session were recorded and summarized by the instructor.

\subsubsection{Sub-session 2: Prototyping and Presentation}
After the brainstorming and discussion, the goal of the second sub-session is to create low-fidelity prototypes so that the previous ideas can be visualized and demonstrated in more concrete forms.
At the beginning of this sub-session, the instructor led a review of the results of the previous session and then informed the participants that they had 30 minutes to create their own prototype using white paper by drawing or handcrafting. Participants were encouraged to discuss their ideas with others and to think aloud during the session, while they still needed to create prototypes individually. After the creation was completed, the participants presented their prototypes to the group and developed open-ended discussions on the prototypes.


\subsubsection{Data Analysis} 
We recorded the entire procedure of the co-design activity with the permissions of the participants, transcribed oral discussions into texts, and collected the prototypes they designed. 
Using the method proposed by Chen and Zhang ~\cite{chen2015remote}, three authors worked as coders to conduct open coding and categorization of the design prototypes. 
In particular, components involved in the design of each prototype (such as expected input and output), information flow, context, and other visual elements (such as present style, text content, and colors) were coded by referring to related work about augmenting presentation and communications ~\cite{liu2023visual, liao2022realitytalk, de2024caption}.
Finally, we revealed qualitative findings regarding how impact captions can support interpersonal communication and derived design goals for a prototype system. 

\subsection{Findings}
Through the co-design activity, we identified three aspects of interpersonal communication in which impact captions can help.

\subsubsection{Enhancing Emotional Expression}
Emotions are important in communication but sometimes it could be implicit and inconspicuous. Impact captions can be a way to materialize human emotions into visible and tangible shapes in social VR. With impact captions, information senders can share their emotional feelings in visible and touchable forms so that the receivers can perceive the sender's moods and intentions with multi-modal perceptions beyond verbal signals and body movements.

\subsubsection{Highlighting Key Information in Speech}
Impact captions can be used to highlight key information in a conversation. Unlike normal captioning system that aims to display all the words in a speech, impact captions should be made for words that conveys key information.
Additionally, typeface, color, and text size are useful dimensions to enhance the visual attractiveness of impact captions to make the key information visible and outstanding in a lengthy speech in VR.


\subsubsection{Identifying Speakers}
Although spatial audio allows users of social VR to trace the source of sounds and identify speakers, it still lacks accuracy and explicit signals, especially when multiple people speak simultaneously ~\cite{yan2023conespeech}.
Impact captions, with their positioning or speech bubble identifiers, can help users to easily identify the source of speech or sound through visual cues. This can be a complement to the audio cues to enrich the communication experience in social VR.


\subsection{Design Goals}
\label{section_design_goals}
Based on the qualitative findings from the co-design activity, we propose the following design goals to implement a proof-of-concept tool to support interpersonal communication in social VR through impact captions.

\subsubsection{G1: Generating Impact Captions in Time}
It was agreed by the participants that the impact captions should be generated when users were speaking, as voice-based ``face-to-face'' conversation is the most common way to interpersonal communication in social VR. Ideally, impact captions should appear simultaneously with the speech.

\subsubsection{G2: Maintaining Readability and Clarity}
Once impact captions were generated, a key aspect in making them useful for communication is to maintain readability and clarity so that other users in the VR space can clearly perceive the information carried out by impact captions. This requires each impact caption to appear in a proper location with a correct facing direction. In addition, multiple impact captions should not overlap each other.

\subsubsection{G3: Enabling Interpersonal Interactions with Playful Impact Captions}
Interactive impact captions could enhance or introduce new forms of interaction between users, making these captions a type of new medium that facilitates connections among multiple users in social VR. 
Taking advantage of the interactivity capabilities provided by VR techniques, users should be able to intuitively handle and play with the impact captions generated by themselves.

\section{SpeechCap: System Design and Implementation}
We implemented \system{} as a proof-of-concept system to demonstrate and evaluate our impact-caption-inspired approach to support interpersonal communication in social VR. According to the design goals (\autoref{section_design_goals}), \system{} takes the user's real-time speech to drive the generation of impact captions in VR (G1), uses rule-based methods to filter out non-important words to be rendered as impact captions in VR (G2), and provides multiple impact-caption-mediated interactions for communication (G2, G3).


\subsection{Technical Pipeline: Algorithms, Software, and Hardware}
\label{sec_pipeline}
\begin{figure*}[!ht]
    \centering
    \includegraphics[width=\linewidth]{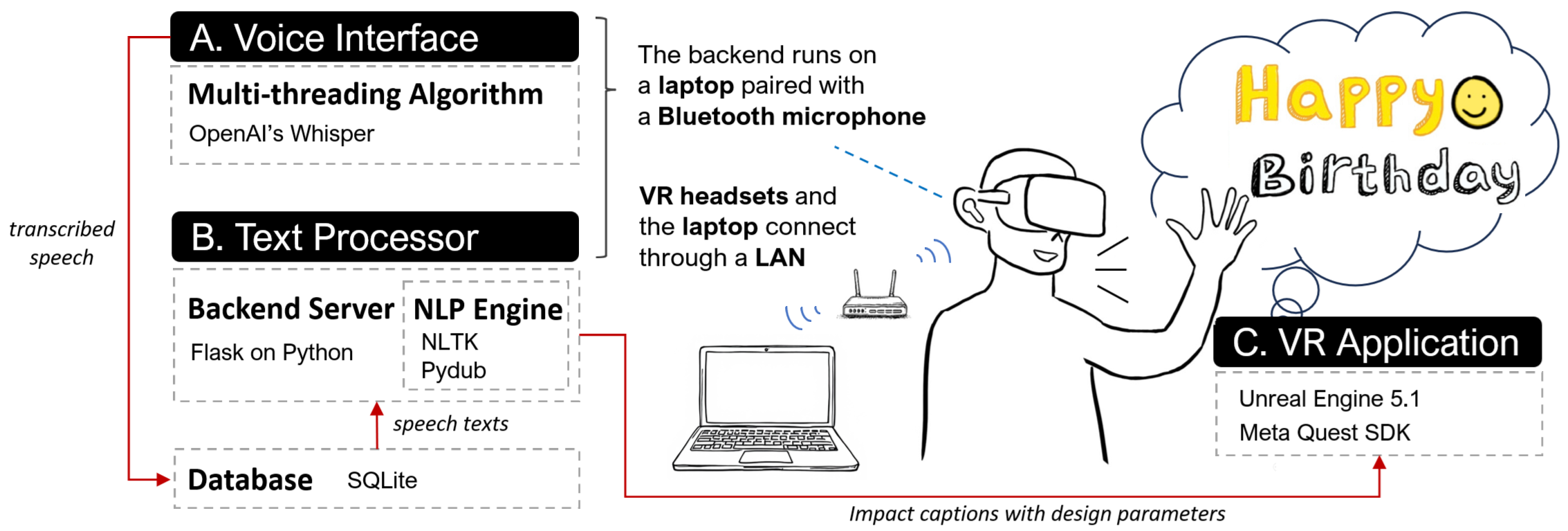}
    \caption{
    \altcolor{
        \system{} System Overview. 
        The system consists of three key modules: (A) \textbf{Voice Interface} that processes real-time voice input and stores the transcribed texts to a shared database,
        (B) \textbf{Text Processor} that distills transcribed texts into impact captions and decides the design of each caption, and 
        (C) \textbf{VR Application} that keeps polling the Text Processor for fetching upcoming impact captions and renders the VR space.
        As for hardware settings, Voice Interface and Text Processor run on a laptop that is paired with a Bluetooth Microphone for collecting voice input. A local area network (LAN) is configured to support the connection between the laptop and VR headsets and among multiple sets of hardware devices for multiple users.
    }}
    \Description{.}
    \label{fig:pipeline}
\end{figure*}

The technical pipeline of \system{} (\autoref{fig:pipeline}) consists of three main software modules namely \textbf{Voice Interface}, \textbf{Text Processor}, and \textbf{VR Application}.
Voice Interface transcribes speech to text in real time and stores the transcribed text for subsequent processing.
Text Processor determines which words should be made as impact captions and decide the appearance of captions based on the analysis of speech texts. VR Application provides VR environment for users to engage in communication through interactive impact captions.

\subsubsection{Voice Interface}
\label{sec_voice_interface}
We implemented a Python module with basic real-time speech-to-text capability using OpenAI's Whisper \footnote{https://github.com/openai/whisper} ~\cite{radford2023robust} as the core model for speech recognition.
As a state-of-the-art AI model designed for speech recognition tasks, Whisper's accuracy is close to that of professional human transcribers ~\cite{radford2023robust}, and a speed exceeding 10 words per second when using the minimal pruned model (i.e., ``tiny.en'') for transcribing English on an ordinary hardware platform ~\cite{haz2023study}.

Since Whisper was not originally designed for real-time transcribing, we employed a custom multi-thread algorithm to process voice input in near real time.
At the start, the algorithm divides the input voice stream into fragments of fixed duration (e.g., 1 second by default). Once a fragment is created, the algorithm would allocate a thread to call Whisper for transcribing, and then save the text result with a sequence number. 
When running the algorithm, the sequence number starts from zero and increases by one each time for a newly processed fragment. The sequence number mechanism ensures that the downstream Text Processor and VR Application can always maintain alignment with the original speech when they process texts and render impact captions in VR.

In terms of accuracy, our method may slightly underperform compared to the core model due to the use of a multi-threading strategy that processes speech chunks sequentially
Thus, a single word might be split into two successive chunks and fail to be recognized, although this is unlikely.
For speeding up, Whisper provides a sort of pruned model with smaller sizes to reduce computational consumptions. 
And in our algorithm, the maximum number of simultaneous working threads is configurable. More available threads can accelerate transcribing while requiring more computational power.
Ideally, the delay between a word being recorded and transcribed can be calculated by summing up the chunk duration (e.g. 1 second), Whisper processing time, and the time of waiting for an available thread. 
Particularly, in our user study, we adopted the ``tiny.en'' model of Whisper (with the minimum number of parameters and specified for English) and a maximum of 4 threads to maintain an acceptable transcribing speed for running the Voice Interface on Apple MacBook Pro with the M1 Pro chipset.

Our speech recognition algorithm is scalable in terms of both accuracy and efficiency. 
By employing Whisper models with different sizes or other SOTA speech recognition models, both dimensions could be improved, although the trade-off between accuracy and efficiency remains. 
Except for the core model, by configuring different chunk durations and the number of available threads (i.e. thread pool size), the efficiency is optimizable.
In addition, advanced hardware resources other than laptops can bring significant improvements to the performance of the Voice Interface module.

\subsubsection{Text Processor}
\label{sec_text_processor}
To decide which words should be made as impact captions and the specific visual design of each individual caption, we developed the Text Processor module to analyze transcribed speech using NLTK \footnote{https://www.nltk.org/} and Pydub \footnote{https://pydub.com/}.
Text Processor and upper-stream Voice Interface are connected by a shared database (SQLite) in which transcribed texts are stored by Voice Interface. And to respond to the needs of the downstream VR Application, we encapsulate the Text Processor in a back-end web server using the Flask framework \footnote{https://flask.palletsprojects.com/}. The server is responsible for providing processed data (i.e., the content and visual design of impact captions) for the VR Application through Web APIs.

For keyword extraction in Text Processor, we followed a strategy using the part-of-speech (POS) tag to filter out function words (e.g., \textit{conjunction}, \textit{particle}, and \textit{adposition}) as these categories of words mainly convey grammar functionalities with less semantics, and to keep words from the tags \textit{noun}, \textit{verb}, \textit{adjective}, and \textit{adverb}. In addition, we also keep the words of \textit{interjections} as they indicate emotional expressions in conversations, and emotional expression is one of the primary usages of impact captions based on our previous findings.

For the visual design of impact captions in VR, Text Processor applies a rule-based approach to generate a customized visual appearance regarding the design space (\autoref{section_design_space}) for each individual impact caption. The rule-based approach utilizes semantic and acoustic features extracted from the user's speech to determine the appearances on each dimension of the design space for an impact caption instance. Details about the rules are discussed in the next section \autoref{section_mapping}.

\subsubsection{VR Application}
\label{sec_vr_application}
We implemented a VR application to render interactive impact captions and provide multi-user virtual reality environments to simulate social VR conversation scenarios. Technically, the VR application is built with Unreal Engine 5.1~\footnote{https://www.unrealengine.com/en-US/unreal-engine-5} and Meta XR Tools~\footnote{https://developers.meta.com/horizon/develop/}, as we adopted Meta Quest 2 as the target VR hardware device.
Besides the overall settings, we utilized the built-in 3D Text Actor of Unreal Engine~\cite{unrealtext2024} to implement the impact captions in VR.

To achieve a multi-user VR environment for at least two players (i.e., running VR simulation in the same virtual session on multiple HMDs), we developed our VR application based on the ``Listen Server'' mode offered by Unreal Engine ~\cite{unrealmultiplayer2024}. In this mode, the first player who starts the application will take the responsibility as a ``server'' to create and host a session over a local area network (LAN), while the ``server'' itself is also a ``client''. Once a session has started, other players in the same LAN can join the session and play as a pure ``client''.
Technically, whether or not to play the role of ``server'', each individual VR HMD in the multi-player session needs to simulate the entire virtual reality environment independently. 
To make the independently simulated virtual worlds have the same appearances for all the players, the ``clients'' have to share the same parameters for rendering objects and simulating interactions through a replication mechanism.
The replication mechanism continuously synchronizes the parameters from the ``server'' to the ``clients'' and reports changes made by the ``clients'' to the ``server''. In this way, multiple players can eventually experience the same VR environment using their own devices.

Nevertheless, the replication mechanism does not force every ``client'' to strictly have the same appearance all the time, since each individual ``client'' always reserves full control of the objects in its own simulation.
This allows us to make local modifications to the objects before they are finally rendered on each individual ``client''.
In this way, we adjust the rotation parameters of the impact captions on each ``client'' HMD, making the impact captions always face towards the HMD's owner player in a readable angle (i.e., the replicas of the ``same'' impact caption are rendered to face different directions in different ``clients'').

\subsubsection{Hardware Settings}
\label{sec_hardware}
Hardware settings include a laptop, a wireless Bluetooth microphone, and a VR head-mounted device with controllers for a single user (\autoref{fig:system}). 
Specifically, we employed an Apple MacBook Pro (with an M1 Pro chipset) laptop for running the Voice Interface and Text Processor modules with the associated database and web server, and a Bluetooth wireless microphone connected to the laptop for collecting the user's voice input. Meta Quest 2 with controllers is used to run the \system{} VR Application. 


To create a stable network environment for serving the multi-player session among the devices, we configured a router to provide a local wireless network with fixed IP addresses for each quest device and the laptops. Additionally, during a session, we use a remote meeting software (e.g., ZOOM) on the laptops to transmit the voices of the users so that they can hear each other in the VR environment.

\subsection{Speech-driven Impact Caption Generation}
\label{section_mapping}
\begin{figure*}[!ht]
    \centering
    \includegraphics[width=\linewidth]{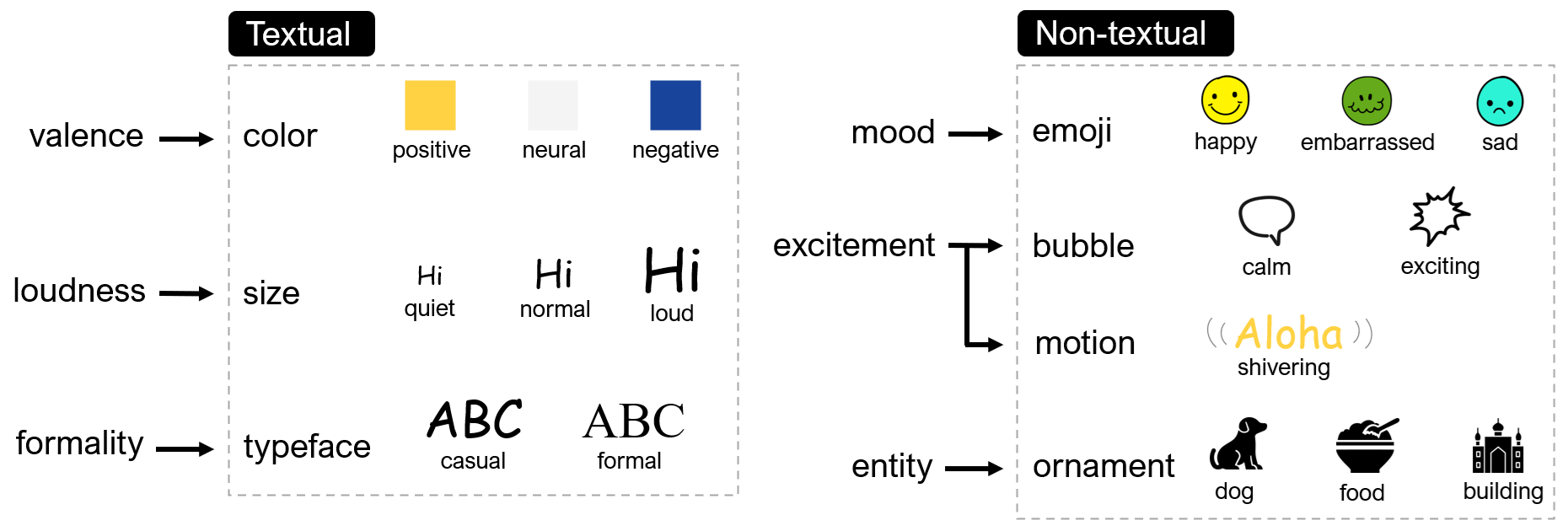}
    \caption{
        Mappings Semantics to Impact Caption Design for Proof of the Concept.
        Valence links to text color where a warm color for positive moods and a cold color for negative moods.
        Loudness links to the size of captions. The larger the louder. 
        Formality links to typeface. ``Time New Roman'' is used for formal and ``Comic Sans'' is used for casual words.
        Emoji is applied for words regarding happy, embarrassed, and sad feelings.
        Speech Bubble and a ``shivering'' motion is applied for words with excitement.
        Ornaments is applied for words representing specific entities.
    }
    \Description{.}
    \label{fig:real_mapping}
\end{figure*}

In \system{}, we used a rule-based method to determine the visual appearance of impact captions across each dimension of the design space.
Our method takes verbal content, affective factors (e.g., valence) and prosodic factors (e.g., loudness) extracted from the speech to compute the visual design of each impact caption (\autoref{fig:real_mapping}), ensuring that the captions align with the context of the conversation.

\subsubsection{Text Color}
Based on our design space (\autoref{sec_space_color}) and the empirical knowledge of color-emotion associations ~\cite{wilms2018color, hanada2018correspondence, plutchik2013theories}, we selected a bright and warm color ({\textcolor[rgb]{1.0,0.82,0.26}{\faSquare}}, close to ``light orange'') for representing captions with positive emotional expressions, a dark and cold color ({\textcolor[rgb]{0.09,0.27,0.61}{\faSquare}}, close to ``dark blue'') for negative captions, and white for neutral captions. 
Technically, the valence of an impact caption is determined by NLTK's ``opinion\_lexicon'', which refers to lists of positive and negative words for sentiment analysis.

\subsubsection{Caption Size}
The size of an impact caption is determined by the loudness (volume) of its belonging speech recording fragment. We estimate the loudness by decibels relative to full scale (dBFS) because dBFS measures volume on the same scale so that comparable results can be generated from different audio recording fragments. Using dBFS, the value 0.0 represents the maximum loudness, and negative numbers are used for lower loudness. For example, -20 in dBFS means that the volume is 20 dBs less than the maximum.
In \system{}, we implemented a rule that allows impact captions to be rendered in three sizes: ``small'', ``medium'', and ``large''. Small size is applied when the dBFS value is under -40. Medium size is applied when the dBFS value is between -40 and -20. Large size applies when the dBFS value is above -20. The louder the speech is, the larger the impact captions will be.

\subsubsection{Typeface}
Referring to the design space and previous studies on human perceptions of typeface ~\cite{bianchi2021emotional, amare2012seeing}, we use typeface to encode and represent the formality of speech. 
As pointed out by previous research, the formality of texts is not a dimension that can be directly calculated but inferred by the words ~\cite{heylighen1999formality}.
Thus, we decided to classify words in speech as either ``formal'' or ``casual'' in terms of their formality and apply a typeface for the two groups, respectively.

To identify the formality of words, we prepared a pre-defined word list of formal words so that we can determine an incoming word by checking if it is in the list.
As for the typefaces, we referred to the literature to guide our design choices.
For formal words, we applied ``Times New Roman'' \footnote{https://en.wikipedia.org/wiki/Times\_New\_Roman}, which is a serif typeface originally designed for serious publications. It is also known to make people feel formal and serious ~\cite{mackiewicz2004people}.
For casual words, we adopted ``Comic Sans'' \footnote{https://en.wikipedia.org/wiki/Comic\_Sans}, as it is also studied by previous research as a casual, sans-serif typeface with a playful and informal look ~\cite{amare2012seeing}. Its visual characteristics feature rounded edges and irregular strokes.


\subsubsection{Emoji and Emoticon}
As a non-textual element that directly links to emotions, we selected three common moods to be explicitly displayed by emojis in our impact captions: a ``smiling face'', a ``sad face'', and an ``embarrassed face'', as faces with common expressions belong to a category of emojis that are known to be commonly used across different cultural backgrounds ~\cite{czkestochowska2022context}.
Similarly to the rules for deciding typefaces, we employed three special word lists for the three moods, respectively. Once a word in the lists is mentioned, the corresponding emoji will be attached aside from the textual part of an impact caption. As the most straightforward example, the word ``happy'' would be accompanied by the ``smiling face'' emoji, while ``crying'' will have a ``sad face'' emoji.

\subsubsection{Ornament}
According to the design space, typographic ornaments include various types of non-alphabetic characters or icons that are relevant to the words within the impact caption. In \system{}, we focus on one of the main purposes of using the ornament, which is to visually highlight objects or entities (e.g., ``cake'') that were mentioned in speech. 
Particularly, we specified a list of keywords of common concepts in daily life for ornaments. The graphic icons for these words are retrieved from Noun Project \footnote{https://thenounproject.com/}, a free open-source icon library. 
The assets of the icons were pre-installed in the VR Application module of \system{}, and the back-end Text Processor will make decisions for each word based on the word list.

\subsubsection{Speech Bubble}
Speech bubble in \system{} is used for two types of words: greetings (e.g., ``hello'') and interjections (e.g., ``HHHHHH'' for laughing), since they indicate attention requirements or speech with excitement.
For impact captions with greeting words, the balloon is rounded-edged to make the captions different from others and attractive. For interjections, a spiky-edged balloon is used to simulate surprising sounds or speech.

\subsubsection{Motion Effects}
A ``shivering'' motion effect will be added to the words with excitement, such as ``shocking'' and ``surprised'', when spawned. 
Rather than adding more motions to the impact captions while they are generated, we leave the potential of triggering motion effects to the users, since VR naturally provides such interactivity over videos where impact captions originate from. In \system{}, the impact captions can always be alive and interactive.

\subsection{Communication with Interactive Impact Captions in VR}
\label{sec_interactivity_functions}
\system{} allows users to have conversations in a shared VR space with the support of interactive impact captions. In \system{}, impact captions are exclusively triggered by speech input. 
\altcolor{
When a particular word (i.e., identified by the rules in Text Processor, \autoref{sec_text_processor}) is mentioned in the speech,
}
\system{} will create a corresponding impact caption at the position roughly in front of the speaker’s avatar at the height of the chest so that users can easily grab the captions. To avoid overlapping, the position will be randomized each time for newly generated captions (\autoref{sec_visual_clutter}).

The visual appearance of each caption is determined by the semantics and affective factors of the corresponding speech (\autoref{section_mapping}). In a conversation, both the speaker and the listener can see and play with the captions, allowing the communication process to be visible, interactive, and playful beyond traditional voice-only experiences.

\begin{figure*}[!ht]
    \centering
    \includegraphics[width=\linewidth]{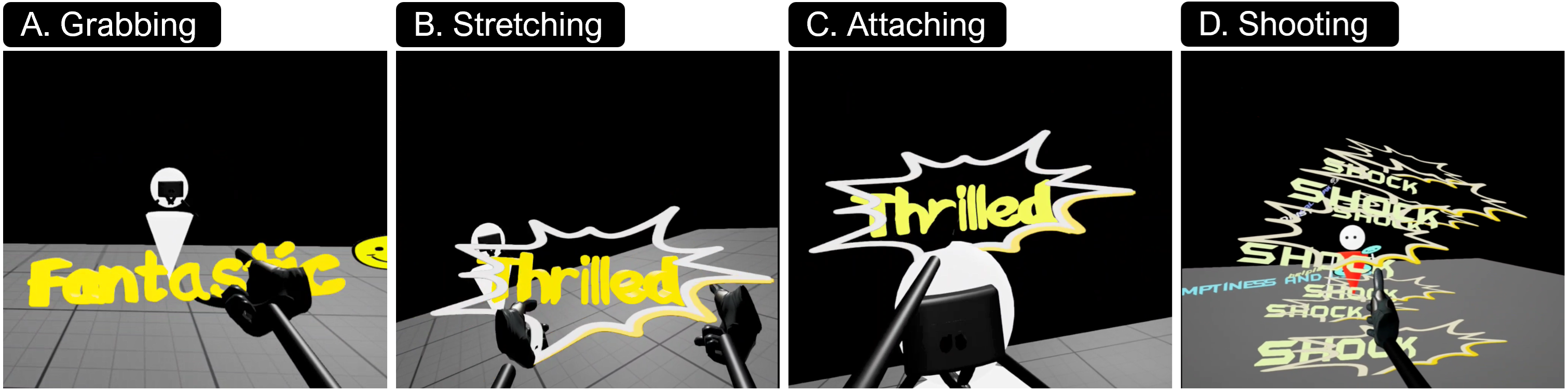}
    \caption{
        Interactions with Impact Caption. 
        (A) \textbf{Grabbing} allows users to hold and place a caption to an arbitrary position;
        (B) \textbf{Stretching} needs two hands to resize an impact caption;
        (C) \textbf{Attaching} allows an impact caption to be attached on the head or body of the virtual avatar;
        (D) \textbf{Shooting} can eject an impact caption forward and trigger the \textbf{Explosion} effect when a collision occurs.
    }
    \Description{.}
    \label{fig:system}
\end{figure*}

\subsubsection{Fundamental Interactions}
\system{} allows users to intuitively interact with impact captions in VR through a set of embodied actions (G3), enabling basic human-caption interactions.
\textbf{Grab} (\autoref{fig:system} A) allows users to hold a caption using VR controllers, adjust its orientation, and move it to a new position, mimicking real-world interactions with a physical item.
When grabbing an impact caption, users can \textbf{Shake} it to trigger shivering and blinking effects.

If both hands grab the same impact caption simultaneously, users can \textit{Stretch} it to resize (\autoref{fig:system} B).
By resizing, users may create huge captions larger than their avatars, or tiny captions like a bullet. 
The ability to adjust size broadens the functionality of impact captions, making them more than just speech-driven subtitles. 
Users can combine captions of different sizes with other actions to enhance interactions with others. For example, making a huge caption and shaking it to trigger the blinking effects to enhance the visual impact to emphasize the key points of conversation.

Furthermore, when holding an impact caption, users can \textbf{Throw} it away by releasing the grab button while moving their arms, just like the natural ``throw'' action. 
As an enhanced version of ``throwing'', users can \textbf{Shoot} a caption, letting it emit towards a direction in a straight line. The shooting action is bonded with the front trigger button on the controllers. Users can hold the button for up to 3 seconds to charge the action. Longer charging time rewards a stronger initial force for emission, resulting in a high initial velocity of the ``bullet'' caption.

\subsubsection{Decorate Avatars for Self Expression}
Avatars play an important role in social VR that supports self-presentation and reflects users' emotions and personalities ~\cite{sykownik2022something, freeman2021hugging}. 
In \system{}, users can \textbf{Attach} an impact caption to their avatar’s head by tapping or touching it while holding the caption (\autoref{fig:system} C). Once attached, the caption remains fixed like a ``tag'' or ``hat''. 
This feature is inspired by the typical usage of impact captions in TV shows where impact captions are used to tag the characters, revealing their emotional status. 
Besides the head, users can also attach an impact caption to a body part of the avatar, causing it to revolve around the avatar periodically and creating a circling motion effect.

\subsubsection{Mediating Interpersonal Interactions}
In \system{}, impact captions can react to user actions and avatars, serving as a medium to facilitate interactions.
When thrown or shot by users, impact captions start flying in the virtual space.
Once a flying impact caption collides with another caption or avatar, it triggers an explosion effect (\autoref{fig:system} D), generating multiple replicas that emit outward like fireworks.
Such a dynamic effect may be used as a non-verbal cue to attract attention and add more fun to the social VR experience.

\subsubsection{Reducing Visual Clutter and Overlapping}
\label{sec_visual_clutter}
To maintain the readability and clarity of floating impact captions in conversations (G2), \system{} introduces a time-to-live (TTL) mechanism to manage the number of captions.
The TTL mechanism allows each caption to live for 5 seconds once generated.
If users do not interact with a newly generated impact caption within the time period, the caption will automatically disappear afterwards. 
To keep an impact caption, users can simply touch or grab it before it disappears.

Once the caption has interactions with a user, it will no longer be constrained by the living time limitation until users intentionally delete it by holding it and clicking a button on the VR controller (i.e., ``X'' button for the left-hand controller and ``A'' button for the right-hand controller).

Additionally, to avoid multiple impact captions overlapping with each other when a user keeps speaking, \system{} randomizes the positions for spawning. Specifically, the exact spawn position of a caption is calculated by a fixed position plus random offsets on $x$, $y$, and $z$ dimensions. This strategy creates a ``word cloud'' effect in front of the speaker’s avatar.

\section{Application Scenarios}
\label{sec_apps}
Besides the scenarios and potential creative use cases committed in the user study, \system{} further supports a variety of communicative needs under different contexts. We briefly introduce the following additional application scenarios to illustrate the generalizability of \system{}.

\subsection{Making Social VR Accessible to Deaf and Hard-of-Hearing (DHH) People}

Recent research revealed that current VR applications fail to provide sufficient accessibility support for deaf and hard-of-hearing (DHH) people to experience immersive digital content or engage in remote communication and socialization ~\cite{jain2021towards, borna2024applications}.
Captioning systems, widely used to enhance accessibility in other digital media such as videos, remain under-explored in VR contexts ~\cite{kim2023visible, de2023visualization, de2024caption}.
\system{} not only has similar functionality to previous systems that convert speech into visible captions for DHH users ~\cite{kim2023visible, de2023visualization, li2022soundvizvr}, but also extends this concept by integrating visual cues and interactivity to convey rich non-verbal information (e.g., valence, excitement).

For example, in a conversation about a vacation plan in social VR with \system{}, a DHH user can discern the speaker’s positive mood through bright-colored captions accompanied by a laughing emoji. A shivering motion further indicates the speaker’s excitement about the trip. 
By dragging and placing impact captions of the names of the must-visit spots mentioned in speech, the speaker can demonstrate the plan with clarity. 
Additionally, when a speaker picks up a caption labeled ``volcano'' while discussing hiking and exploration, the corresponding icon reinforces the speech context.
Overall, informative captions, engaging visuals, and interactive features in \system{} expand opportunities for DHH individuals to connect with others and engage with the virtual world in social VR

\subsection{Enhancing Interactions for Teaching and Learning in Social VR}
VR has shown great potential in education by enhancing social presence in remote learning and providing access to learning contexts otherwise unavailable in reality ~\cite{thanyadit2022xr, peng2021exploring, jensen2018review}. However, it remains far from mainstream adoption due to challenges in creating easily reviewable educational content, ensuring inclusivity, and supporting collaborative learning ~\cite{jin2022will}.   
In virtual lectures, \system{} enables instructors to create instructional cues by placing keyword-based impact captions in the virtual space. Shared interactive captions facilitate interaction between instructors and students, supporting activities like Q\&A and hands-on demonstrations.

For example, when teaching the structures of plants in a botany class, the instructor would usually introduce the roots, stems, leaves, and flowers in sequence. With \system{}, once these words were mentioned, the relevant impact captions would be generated with texts accompanied by symbolic icons.
Then, the instructor can grab and stretch to enlarge the caption for emphasizing, and further captures the students' attention by shaking the caption to trigger the blinking effect.
If a student has questions about a specific concept mentioned before, he could also make an impact caption of words for that concept and shoot the words, creating an explosion of the words so that the instructor can easily know which part the student is confused about. 
In addition, further use cases can be explored to adapt to different contexts in teaching and learning. \system{} in the current stage provides a starting point to a way to achieve engaging and playful interpersonal interactions for social VR users in the future.

\subsection{Facilitating Engaging Live Streaming Experiences in Social VR}
Recent research suggests that VR streamers face challenges in building emotional connections, as VR headsets obscure facial expressions, making it harder for viewers to perceive their emotions directly ~\cite{wu2023interactions}, while emotional connection is a key factor in live streaming ~\cite{lu2018you}.
\system{} visualizes emotions for streamers by detecting moods in speech and adding appropriate emojis and motion effects to captions, and it also enhances communication by rendering viewers’ reactions as interactive captions in the streamer’s virtual space, fostering engagement.
These features of \system{} address the need for visible and spatialized objects, creating more engaging live streaming experiences ~\cite{wu2023interactions, lu2018you}.

For example, in a typical VR live streaming scenario, a streamer was self-evaluating his playing performance on Beat Saber, a popular VR rhythm game. He didn't do well in the game but would like to tell the audience that he had a lot of fun playing it and felt like a champion. This strong emotion can be communicated and augmented by \system{} with colored captions of words like ``Fun'' and ``Exciting'' accompanied by laughing emojis. 
He then made a ``Champion'' caption with a trophy icon and attached it to his avatar's head like a funny crown to show self-mockery.
By shooting the caption ``Exciting'' outwards, the explosion motion effect can be triggered once the caption crashes with an object in the VR scene, making the atmosphere more interesting and engaging. From the viewers' perspective, they can also send their feedback and comments as impact captions, using the visible and interactive captions to make humorous responses to the streamer's poor performance.


\section{User Study}
To understand the effectiveness of the design space and \system{} system in supporting interpersonal communication in social VR, we conducted a user study, inviting 14 participants of various backgrounds from local universities to experience simulated conversations in VR with interactive impact captions using \system{}. User feedback was evaluated from both quantitative and qualitative perspectives through a post-study survey and semi-structured interviews. 


\subsection{Participants}
We recruited 14 participants (6 female and 8 male, aged between 18 and 35) from local universities through advertisements on social media. 
None of the participants overlapped with the four experts involved in the previous co-design activity. 
Five participants (P1 to P5) were novice users who had never used VR head-mounted displays (HMDs) before, while the other nine participants (P6 to P14) self-reported having extensive experience with VR games, social VR, and application development.

\subsection{Procedure}
The user study was conducted in a one-on-one format, where an experimenter (either the first or second author of this paper) guided each participant through the procedure.
The study sessions lasted 75 to 90 minutes and were held in a university lab space. After the study, the participant received a gift card valued \$50 in local currency. For analysis, the study processes were fully audio-taped and the activities in VR were screen-recorded.
Each session utilized two Meta Quest 2 devices and two laptops, each paired with a Bluetooth microphone, for the experimenter and the participant. An additional router was configured to host a local area network to serve the multi-user session of \system{}'s VR application (\autoref{sec_hardware}). 


The primary goal of this user study was to evaluate \system{} using two common social VR scenarios: (1) a private conversation between two friends and (2) a public discussion on a debatable topic.
The dyad conversation scenario simulated a private conversation designed to evoke rich emotional expressions.
In this scenario, the experimenter and the participant play as close friends to talk about their daily life and personal feelings with the help of impact captions. 
In the multi-person conversation scenario, the participant joined an ongoing discussion, listened to others, and then shared their own opinion.
In the two scenarios, participants will have opportunities to play the role of both speaker (i.e., information sender) and listener (i.e., information receiver) in turns, allowing them to play with various impact caption instances in very different visual and interactive appearances.
Specifically, a full user study session consists of four sections as follows: 

\subsubsection{Introduction and System Walk-through (\textasciitilde20 min)}
At the beginning, the experimenter briefly introduced the background and main purpose of \system{} (i.e., facilitating interpersonal communication in social VR). Next, the experimenter guided the participant to get familiar with basic usages of the hardware devices if they were not, and then walked through the \system{} system. During the walk-through, the experimenter provided verbal and hand-in-hand guidance, ensuring that the participant could see and generate impact captions with their own speech, and know how to interact with the impact captions in VR. After the walk-through, the experimenter would inform the participants about the next dyad and multi-person scenarios.

\subsubsection{Dyad Conversation Scenario (\textasciitilde15-20 min)}
Once the participant and the hardware settings were ready, the experimenter would take another set of devices, go to a separate room (avoiding conflict with the speech-to-text program running on the laptop), and initiate the conversation in VR by telling a pre-defined sad story about losing her cat recently. During the storytelling, the experimenter employed several impact captions to enhance the speech and seek emotional support from the participant. The participant was then asked to comfort the experimenter by telling some happy moments from recent daily life, using impact captions generated along with the conversation's development. 

\subsubsection{Public Discussion Scenario (\textasciitilde15-20 min)}
In this scenario, the participant was asked to join a discussion of the topic ``\textit{what privacy and security issues will arise around AI?}''
The discussion was started between the experimenter and a ``NPC'' character who was configured to speak based on a pre-defined script and timeline, due to the limitation of hardware resources. Once joining the session, the participant was asked to first listen to the ongoing conversation for a while and then present her/his own opinions with the help of impact captions.

\subsubsection{Survey and Semi-structured Interview (\textasciitilde25-30 min)}
After the two conversation scenarios, the participant was asked to complete a survey about their experience with \system{}. The participant was also encouraged to think aloud while completing the questionnaire. When the survey was done, the experimenter would conduct a semi-structured interview with the participant, diving into their experiences and thoughts in depth.

\subsection{Results and Findings of Post-study Survey}
The post-study survey questionnaires consisted of two sections. The first section employs four questions that ask whether the impact caption design space is understandable, interesting, and meaningful.
The second section contains six questions about the user experience with \system{}, considering enjoyment, emotional expression, clarity of presentation, and overall experience of having conversations with impact captions in VR. The questions are designed in a 7-point Likert scale format ~\cite{joshi2015likert}, requiring the participants to respond with their level of agreement on the statements.

\begin{figure}[htb]
    \includegraphics[width=\linewidth]{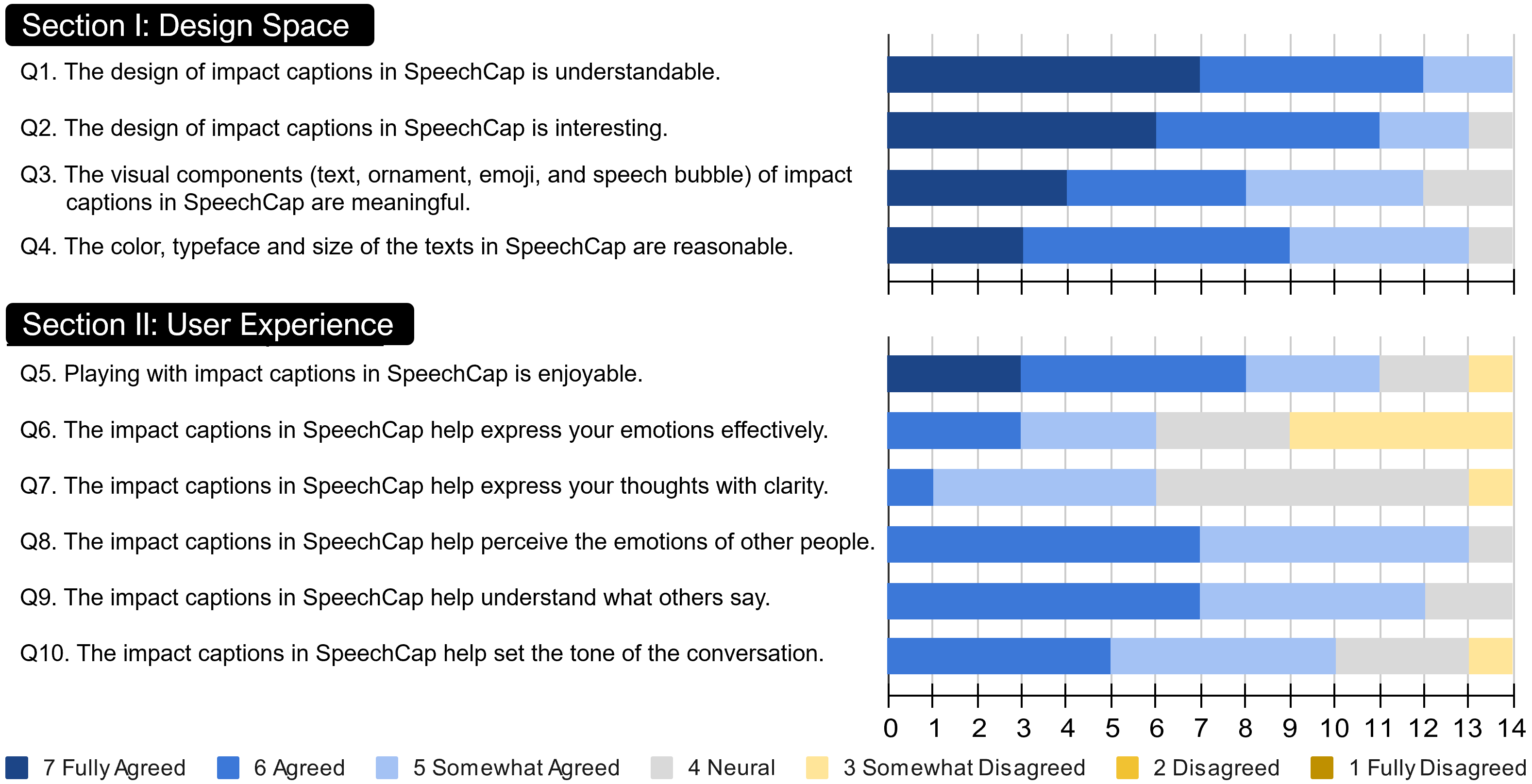}
    \caption{
        Post-study Survey with Results. 
        The survey consists of two sections in which the first section contains four questions (Q1-Q4) about the design space of impact captions and the second section contains six questions (Q6-Q10) about the experience with SpeechCap. 
        Overall, participants believed the design space of impcat captions was meaningful and the experiences of using \system{} is enjoyable and helpful.
    }
    \Description{.}
    \label{fig:survey}
\end{figure}

In the first section (\autoref{fig:survey} upper), participants agreed that the design of impact captions for VR communication was both understandable (Q1, M=6.36, SD=0.74) and interesting (Q2, M=6.14, SD=0.95). 
Regarding the visual design of individual captions, 12 participants agreed that combining non-textual elements (ornaments, emojis, speech bubbles) with text was meaningful (Q3, M=5.71, SD=1.07). 
And 13 participants believed that the visual design (i.e., color, typeface, and size) of textual elements was reasonable (Q4, M=5.79, SD=0.89). No participants responded with disagreement on the design of impact captions.

In the second section (\autoref{fig:survey} lower), most of the participants (11/14) agreed that their experiences with \system{} were somewhat enjoyable (Q5, M=5.50, SD=1.22). 

\altcolor{
From an information sender's perspective, participants diverged on the effectiveness of using impact captions for emotional expression in social VR (Q6, M=4.29, SD=1.20). Six participants agreed that impact captions can help express their emotions effectively, while the other five committed objections. Two participants responded with a neutral opinion. 
For the next question about clarity (Q7, M=4.43, SD=0.76), six participants responded with agreement, while five participants stayed neutral and one responded ``somewhat disagreed''.
The results of these two questions indicate that \system{} is not always helpful for users to express their emotions and thoughts. 
}

From a receiver's perspective, the majority of the participants (13/14) agreed that they could perceive the emotions of others by seeing and playing with the impact captions in VR (Q8, M=5.43, SD=0.65). 
\altcolor{
Most participants (12/14) agreed that impact captions improved their understanding of others in VR (Q9, M=5.36, SD=0.74). 
}
In addition, ten participants agreed that impact captions were useful in setting the tone of conversation (Q10, M=5.00, SD=0.96).
\altcolor{
Overall, \system{} seems more helpful for information receivers as they felt the conversation clear and understandable, while around half of information senders thought the system failed to improve their expressions.
}

\subsection{Results and Findings of Semi-structured Interviews}
During the study, All participants completed both scenarios, playing the roles of both sender and receiver in conversations. Creative usages of impact captions in VR emerged from their play-through. 
In the semi-structured interviews, we discussed questions regarding the impact captions' design space and user experiences of \system{} with the participants based on their individual performance.
Overall, participants responded positively, describing the impact captions as ``interesting and enjoyable'' (P1, P2, P4, P10, P13).

Two authors analyzed the transcribed scripts of the interviews by iterative open coding ~\cite{corbin2014basics} and finally revealed the following findings showing the meaningfulness of our impact captions' design space, and the usefulness of \system{} system.

\subsubsection{F1: Textual Elements Can Effectively Support Conversations in Social VR}
\label{finding_text}
Participants generally agreed that visible impact captions enhanced their conversational experience in VR (P2, P3, P4, P6, P7, P11, P12, P13, P14). 
On one hand, text elements help users accurately express complex and profound meanings. P12 emphasized the expressive capability of impact captions: ``\textit{The texts within impact captions have a high potential in expressiveness, because texts can convey complex, abstract concepts as well as accurately represent ironic or literary content, which non-textual elements cannot achieve.}''

On the other hand, some participants (P2, P7, P8, P13, P14) believed impact captions help extract and emphasize key information from dialogues. 
In \system{}, impact captions automatically disappear if not touched by users in several seconds (\autoref{sec_visual_clutter}). Therefore, besides filtering trivial words in the back-end processing (\autoref{sec_text_processor}), \system{} also allows users to actively select impact captions with keywords they consider important to remain in VR. ``\textit{...if I was sharing some knowledge with others with \system{}, I could easily make the keywords of my speech persistent and let them floating around my avatar so that my audiences won't lose focus...}'', said P7. 
And when talking about the multi-person conversation scenario, P14 said ``\textit{I'd appreciate \system{} if it can always correctly recognize and show the meaningful words by impact captions. This prevents me from listening to lengthy speeches in meetings.}''

Additionally, one participant (P7), a non-native English speaker, felt that impact captions lower the barrier for her to engage in English conversations. P7 said, ``\textit{English is not my native language, and when speaking with native speakers, I often struggle to understand due to different accents and fast speed of speech... but impact captions allow me to clearly and accurately see what others are saying, and they can remain in the VR space as a conversation record for later review.}''

\subsubsection{\altcolor{F2: Non-textual Elements are Engaging but Ambiguous}}
\label{finding_ambiguity}
Overall, participants agreed that the non-textual visual elements in impact captions are engaging and making the speech-driven captions really ``impactful'' (P1, P2, P3, P4, P5, P6, P8, P10, P11, P12, P13). 
P1 said ``\textit{I love the idea you combine those cute emojis with the colorful texts. It makes those captions not only interesting but also meaningful.}'' 
Among the multiple non-textual dimensions in the design space, the colors, ornaments, emojis, and motion effects are perceived by participants as the most noticeable features (P1, P3, P4, P7, P10, P11, P12). 

\altcolor{
Nevertheless, the non-textual elements in impact captions cannot convey meanings as clearly as texts can do. The ambiguity was reported across multiple dimensions, including color and emoji.
}
First, the current colors of impact captions in \system{} were reported to be confusing and ambiguous by some participants (P1, P8, P9). P1 asked ``\textit{I noticed that those impact captions had different colors, but actually I didn't figure out why a word should be rendered in blue while others were not?}'', and P9 made another similar comment saying that ``\textit{I don't quite understand why you choose orange color for the word "happy" while other words are in white.}''. 
Similar issues arise for emojis. P4 commented, ``\textit{Expressing with visual elements alone can lead to misunderstandings; some people like to use a smiling face for implicit sarcasm or be \'ironic\'}.
These cases may relate to findings of previous research that different people may have different perceptions of the same visual elements when conveying non-verbal meanings ~\cite{hanada2018correspondence, wilms2018color, czkestochowska2022context, miller2017understanding}, revealing the existence of ambiguity in our design of impact captions.

However, on the other hand, three participants (P3, P10, P12) believed that the integration of textual and non-textual elements in impact captions can also mitigate the potential ambiguity. P3 highlighted, ``\textit{I might misunderstand the meaning when only seeing the smiling face, but if there was a word "happy" nearby, it would be much clearer,}''. In other words, the meanings of different components of a single impact caption can mutually reinforce each other, reducing the chances of misinterpretation by the recipient. 
When discussing the color-mapping rules in \system{}, P10 commented, ``\textit{I think using different colors to show the emotions in a conversation is a good idea, and I don't worry about misunderstandings because when the overall tone is positive, I can see most words are in orange. The number of colored captions not only can avoid the ambiguity produced by a single word, but also can reflect how strong the emotional expression is.}''

\altcolor{
In summary, the ambiguity of non-textual elements could reduce the clarity in conversations, while it is still possible to deal with the unavoidable ambiguity with design considerations, such as using the combination of multiple dimensions to complement each other.
}

\subsubsection{F3: Interactivity is a Key to Engaging Communication Experience in Social VR}
\label{finding_interaction}
Most participants (P1, P2, P4, P5, P6, P7, P8, P10, P11, P13, P14) agreed that the interactivity of impact captions in \system{} could bring novel collaborative social experiences and help develop a sense of intimacy between the users in social VR. ``\textit{Throwing a caption to people in social VR makes me think of the "squat and strait" movement before starting a game when playing "Fall Guys", in which everybody feels fun together.}'', said P6. She believed the interpersonal actions mediated by impact captions in \system{} were meaningful for building connections with others in social VR, even such interactive capabilities were somehow ``\textit{unnecessary}'' for a speech-driven captioning system. 

Besides the fundamental interactions, more creative usages of impact captions were explored by the participants.
Note-taking is one method commonly proposed. P7 recalled his experience of the multi-person discussion scenario in our study and said, ``\textit{During the group discussion, I would like to grab the keywords mentioned by other members, enlarge them, and place them in the virtual space, just like using a whiteboard,}''. P4 also talked about similar use cases, saying that ``\textit{I used to capture the key points mentioned by others and record them with notes in meetings. In \system{}, impact captions can be the notes, I can just take the captions aside ... this could also be used in multiplayer reasoning games, right? You can note what others have said (as a basis for reasoning),}''.
Note-taking allows impact captions to persistently present the conversation in the virtual space like dialogue records, achieving a means of asynchronous communication. ``\textit{Impact captions can be like email ... other users can read the meeting record whenever they join the session, as long as the impact captions remain there.} said P12.

Another creative use case was conducted by P10. When conducting the dyad conversation scenario in which there was a sad story about a missing cat, he purposefully generated several impact captions with negative words and used these captions as blocks to construct a small shelter in VR by resizing and placing. P10 explained, ``\textit{When I felt sad and uncomfortable, I always find somewhere to hide myself avoiding meeting with other people. So I take these many "sad" words to build a blue room for taking a break.}''. He also invited the researcher to join his room made by impact captions to relax in VR.

\subsubsection{F4: Needs for Identifying Keywords in Conversations}
\label{finding_keywords}
\altcolor{
Other drawbacks of \system{} were exposed by the user study, too.
}
One of the most commonly asked questions by the participants when they started using \system{} was about how we decided the words to be rendered as impact captions, as they usually found there were too many impact captions emerging when speaking continuously.
After knowing our simple POS-tag-based filtering method to remove functional words, most participants suggested an improvement on it. 
P2 suggested, ``\textit{Don not turn every word into an impact caption. Please extract the key information (of the speech) and only show impact captions of the relevant keywords.}'' 
Other participants, like P11, thought too much text content increased the reading and comprehension costs and then reduced the overall experience of having a conversation with \system{}. P11 complained, ``\textit{No one wants to read that much text when using VR.}'' 
Particularly, in the multi-person conversation scenario, as more impact captions were spawned and accumulated in limited space, some participants (P1, P2) reported that they felt the field of vision was partially obstructed.



\section{Discussion}
The user study findings demonstrate the effectiveness of our impact caption design space and the usefulness of our \system{} system. 
Related to findings and the literature, we first discuss the potentials of using impact captions as a form of visual rhetoric with interactivity to enhance communication and interactions in social VR.
Next, we present an in-depth understanding of the ambiguity in \system{} and proposed future research directions regarding it for designing communication tools in social VR.
Then, we provide three design implications for future tools for interpersonal communication in social VR based on the study results.
Finally, we explain the scalability of \system{}, showing the limitations and corresponding future directions to improve our \system{}.



\subsection{Designing Interactive Visual Rhetoric for Interpersonal Communication in Social VR}
As suggested by previous research, mediating interpersonal communication in virtual spaces involves supporting people in building cognitive and emotional connections ~\cite{mcveigh2021case, palmer1995interpersonal} and exploring self-identity beyond the real world ~\cite{freeman2021body, sykownik2022something, maloney2020talking}. 
However, most of the known communication methods in social VR are still imperfect ~\cite{wei2022communication, dzardanova2022virtual, akselrad2023body, sykownik2023vr, tanenbaum2020make, baker2021avatar}.

\subsubsection{Highlighting the Visual Rhetoric and Interactivity}
To fill this gap, our work incorporates speech-driven captioning mechanisms into social VR with improvements on captions' visual design and interactivity.
Inspired by the original impact captions used in entertaining TV shows ~\cite{sasamoto2014impact}, we designed and implemented a visual rhetoric that integrates textual and non-textual elements to convey verbal and non-verbal information simultaneously, while enabling various human-caption interactions to support speech conversations in social VR

\altcolor{
The user study revealed that our impact-caption-mediated approach have the potential to effectively support verbal communication in social VR (\autoref{finding_text}).
The rich interactivity provided by impact captions fosters more engaging and dynamic interpersonal interactions (\autoref{finding_interaction}).
However, it also faces challenges like the ambiguous non-verbal elements (\autoref{finding_ambiguity}) and the difficulties on keyword extraction (\autoref{finding_keywords}).
Overall, \system{} provides a concrete example of how interactive and expressive visual objects can mediate communication and enhance social interactions in social VR.
}

\subsubsection{Considering Generalizability}
As for the generalizability of our impact-caption-mediated method, the aforementioned application scenarios (\autoref{sec_apps}) demonstrate that \system{} have the potential to be widely applied in different contexts. It can enhance emotional expressions to create engaging and intimate atmospheres, support live presentations for teaching scientific knowledge in educational scenarios, and further provide a promising captioning mechanism that can bring immersive social VR experiences to deaf and hard-of-hearing (DHH) individuals.

\subsubsection{Future Directions}
In summary, our work emphasizes the value of visual rhetoric integrating with interactivity for facilitating communication in social VR, showing how invisible verbal and non-verbal information can be made perceptible and playable. \system{} expands on the ``superpowers'' of social VR ~\cite{mcveigh2022beyond, mcveigh2021case}, leading the way towards unique immersive communication experiences and fostering novel interactions not only between two human users, but also between humans and the digital media.

As the design space of impact captions in \system{} is generalizable, future work can borrow from or build on our impact captions to create novel digital mediums to support communication needs in a wide range of social VR applications, including facilitating distinctive activities ~\cite{chen2024drink, maloney2020falling}, enhancing self-presentations ~\cite{freeman2021body, sykownik2022something, maloney2020talking}, and constructing social connections ~\cite{li2019measuring}. Besides, further studies are required to investigate the long-term influence of using \system{} for communication on users and explore whether such a method can interfere with the ``proteus effect'' in social VR ~\cite{maloney2020talking} to reveal opportunities for expanding and enriching the application scenarios.

\subsection{Understanding the Ambiguity in \system{}}
Ambiguity in computer-mediated communication (CMC) systems is unavoidable, arising from socio-technical factors such as cultural differences, personal preferences, and technical limitations ~\cite{stacey2003against, aoki2005making}.
Impact captions in \system{} also face risks of miscommunication due to inherent ambiguity in the design space.

The user study revealed participants' concerns that the current heuristic rule-based method for determining visual appearances might not meet the diverse needs of users, as individuals may perceive the same impact captions differently (\autoref{finding_ambiguity}).
Previous research pointed out that visual elements in the design space, such as color-emotion association ~\cite{hanada2018correspondence, wilms2018color} and emoji ~\cite{czkestochowska2022context, miller2017understanding}, could imply ambiguous interpretations by different people, which also supports our study findings.

However, inevitable ambiguity in digital media design can be a double-edged sword, presenting both challenges and opportunities. 
Although it may introduce risks of miscommunication, ambiguity can foster creativity ~\cite{stacey2003against}, support negotiation ~\cite{gaver2003ambiguity}, and even enhance the human-like qualities of AI agents in human-AI communication ~\cite{liu2024let}.
Therefore, a trade-off arises: for CMC tools, designers must balance the clarity of communication while maintaining certain ambiguity to meet other design goals ~\cite{stacey2003against, aoki2005making}. This trade-off leads to two directions of dealing with the ambiguity: mitigating and utilizing.

Yet, the way to mitigate ambiguity for communication in social VR has not been fully explored. Though previous work revealed the potential of using biosignal visualizations ~\cite{lee2022understanding}.
As a complement, our impact caption design space and \system{} illustrate another way to reduce the risks of miscommunication caused by the ambiguity in social VR conversations. Particularly, our impact captions can reduce ambiguity by providing contextual information, which is a known effective method for communication in contexts other than social VR ~\cite{cottone2009solving, dey2005designing, miller2017understanding}.

For a single impact caption in \system{}, the integration of textual and non-textual visual elements allows the elements to complement each other and provide mutual confirmation in terms of their meanings. This can reduce ambiguity conveyed by a single element (i.e., a single emoji) (\autoref{finding_ambiguity}). 
At the speech level, the speech-driven approach produces impact captions sequentially aligning with the voice input, in which the conversation contexts could be naturally visualized and presented. 
Moreover, users can persistently present an impact caption by simply dragging and placing it at an arbitrary position in the virtual space (\autoref{sec_visual_clutter}) to intentionally create environmental cues ~\cite{cottone2009solving} for providing contextual information with flexibility.
Overall, these features embedded in \system{} pave the way towards designing communication tools in social VR with the consideration of mitigating ambiguity. 
Future work can refer to the self-explanatory visual design of impact captions to construct visual cues that can convey meanings with clarity and can also leave space for users to actively create environmental cues in the virtual space to support their communication activities.

As for utilizing the ambiguity for particular design aims, which is not covered in the current stage of \system{} and the design space of impact captions, we believe it would be a promising research direction for future work that aims to support interpersonal communication and social interactions in social VR.
With the unique affordances ~\cite{mcveigh2022beyond, freeman2021body, freeman2021hugging, wei2022communication} offered by social VR and growing application scenarios with communication needs ~\cite{maloney2020falling, chen2024drink, mei2021cakevr}, ambiguity should have great potential to foster valuable design and mediums to satisfy specific user needs ~\cite{stacey2003against, gaver2003ambiguity}.

\subsection{Design Implications}
Based on \system{} and the findings, we identify the following potential design implications for designing digital mediums and tools to support interpersonal communication in social VR.
With the fast evolution of VR and related computing technologies, the contexts of conducting interpersonal communication in VR may change rapidly. However, we hope these implications can highlight the invariant knowledge learned from \system{} to inspire future work.

\subsubsection{Presenting Verbal Information Selectively}
Verbal information is fundamental in interpersonal communication ~\cite{liu2023visual}. However, when enhancing verbal information in social VR, it is essential to focus on selective presentation rather than displaying every spoken word.
In our user study, participants suggested avoiding presenting every word in speech, since too many insignificant words could be distracting and cause unexpected visual clutter (\autoref{finding_keywords}). 
Although we have already designed a mechanism to softly control the number of impact captions, trying to minimize the possibility of visual clutter (\autoref{sec_visual_clutter}), the POS-tag-based filtering mechanism in the text processor of \system{} is too simple to be an effective keyword extraction method.
With the fact evolution of AI and NLP technologies, future work can explore the ways to involve advanced NLP technologies such as Large Language Models (LLMs) ~\cite{maragheh2023llm} to achieve keyword extraction with accurate understandings of the speech content and the user's intentions.


\subsubsection{Using Visual Rhetoric to Present Non-verbal Information}
Non-verbal information plays an important role in communication, social interactions ~\cite{mcveigh2022beyond, maloney2020talking, aburumman2022nonverbal, liebman2016s}, and self-presentations ~\cite{freeman2021body, sykownik2022something, zhang2022s} in social VR.
Our work demonstrates that rich visual elements can effectively convey non-verbal information in social VR, highlighting the value of visual rhetoric as a design strategy for communication tools.
However, we currently only use simple mapping rules to decide how captions correspond to specific visual elements in \system{}. These rules are inadequate for sufficiently and accurately expressing the various types of non-verbal information, as pointed out by participants in the user study.
Therefore, future research can start from seeking in-depth understandings of the associations between specific visual elements and non-verbal information (such as color-emotion association ~\cite{hanada2018correspondence, wilms2018color}), and create novel visual rhetoric designs with effectiveness and elegance for supporting various applications of social VR.

\subsubsection{Enhancing the Interactivity of Communication Medium}
The impact captions in \system{} embed rich interactivity, enabling users to create multiple new forms of interpersonal interactive actions (\autoref{sec_interactivity_functions}).
This again demonstrates the unique interactive and immersive nature of social VR ~\cite{maloney2020talking}, highlighting the value of the interactivity for supporting creative, engaging, and emotional communication experiences.
On top of \system{} and previous work, future research can explore more on the interactivity afforded by the digital mediums in social VR, investigating their influences and creating novel designs and technologies to enhance the existing communication channels or even introduce new channels to support interpersonal communication and interactions in social VR.

\subsection{Technical Scalability and Future Improvements}
\system{} is theoretically scalable in terms of architecture and algorithms (\autoref{sec_pipeline}). However, the current implementation of \system{} as a proof-of-concept system has limited scalability when considering real-world deployment. We identified three main aspects of the limitations and discuss possible solutions for future improvements.

\subsubsection{Seeking for Better Speech-to-text Solution}
First, our custom speech transcribing algorithm introduces limitations.
The algorithm integrates the Whisper model ~\cite{radford2023robust} with multi-thread processing for approximate real-time processing, which is inferior to the cloud-based end-to-end speech-to-text services provided by business companies in terms of performance (i.e., speed and accuracy). 
Running on laptops further restricts its performance due to the limited computational resources.
For future improvement, we can replace the voice interface module with a well-developed real-time speech transcribing solution, such as a cloud-based business solution or a locally-deployable SDK.

\subsubsection{Reducing Network Load by Simplifying Architecture}
Second, the architecture of \system{} heavily relies on network communication.
The back-end web server works as a centric hub that generates parameters of impact captions and exposes APIs to feed VR applications.
The VR application, as the user end, keeps polling the back-end server to continuously fetch newly generated impact captions through the network. In the meantime, multiple VR application instances also continuously sync with each other to achieve the multi-user session across the network.
As pointed out by recent research, the growing number of users will significantly increase the network load for social VR applications ~\cite{cheng2022we}. This challenge also applies to \system{}.
For improvement, through the perspective of architecture, the back-end server could be reduced, because ideally, all the computations, including speech transcribing and caption generation, can be processed in place in a single computational node (i.e., the VR HMDs). Though this relies on the growth of computational power provided by advanced chips in the future. 
Without the back-end server, only the synchronization mechanism among VR devices requires network resources.

\subsubsection{Simplifying the Hardware Requirements}
Third, the hardware setting is too complex. Currently, we employ a laptop, a microphone, and a VR HMD to serve a single user. This setting requires coordination among the devices to smoothly run \system{}. Even though the setting is manageable under lab settings for user study, it is far from being used by ordinary people out of the lab.
This limitation can also be addressed with the simplification of architecture, as the laptop and external microphone are used for running the back end for speech input and processing. Once those computational tasks are embedded into VR HMDs, no extra hardware will be needed for \system{}.

\subsection{\altcolor{Limitations and Future Work}}
\altcolor{
Although \system{} demonstrates the potential of impact captions to enhance interpersonal communication in social VR, several limitations remain. 
First, the current rule-based method for determining visual appearances of captions is simplistic and may lead to ambiguity, as users interpret visual elements like colors and emojis differently. 
Future work could incorporate advanced NLP techniques, such as LLMs, to extract contextually meaningful keywords and adapt visual designs dynamically to user preferences. 
Second, the system's reliance on multiple hardware components (e.g., laptops, external microphones, and VR headsets) limits scalability and real-world deployment. Future improvements could focus on integrating all functionalities into standalone VR headsets to simplify the setup. 
Finally, while the study explored dyad and group conversation scenarios, further research is needed to evaluate \system{} in diverse social VR applications, such as education, live streaming, and accessibility for deaf and hard-of-hearing users. Long-term studies are also required to investigate the impact of using impact captions on communication behaviors and social dynamics in virtual environments.
}

\section{Conclusion}
This research highlights the potential of impact captions as an innovative medium for enhancing interpersonal communication in social VR. By seamlessly integrating verbal and non-verbal cues through dynamic and interactive typographic elements, impact captions transform real-time conversations into playful and engaging experiences. Our proof-of-concept system, \system{}, demonstrates the practicality and effectiveness of this approach, while also revealing opportunities for improvement. Specifically, future work could focus on reducing ambiguity in visual designs, improving adaptability for diverse user preferences, and expanding accessibility features to ensure inclusiveness for a wide range of social VR users.



\bibliographystyle{ACM-Reference-Format}
\bibliography{main}



\end{document}